\documentclass[12pt]{iopart}

\usepackage{graphicx}
\usepackage{epsfig}
\usepackage{verbatim}
\usepackage{amssymb}
\usepackage{bm}
\usepackage[T1]{fontenc}
\usepackage{eufrak}

\newcommand{\nn}{\nonumber}
\newcommand{\ind}[1]{{\mathrm{#1}}}
\newcommand{\half}{\frac{1}{2}}

\newlength{\bilderlength}

\newlength{\figsize}
\begin{document}

\title[MRH of one-dimensional interfaces : from Rayleigh to Airy distribution]{Maximum relative height of one-dimensional interfaces : from Rayleigh to Airy distribution}
\author{Joachim Rambeau and Gr\'egory Schehr}  
\address{Laboratoire de Physique Th\'eorique (UMR du
  CNRS 8627), Universit\'e de Paris-Sud, 91405 Orsay Cedex,
  France}

\date{\small\today}
\begin{abstract}
We introduce an alternative definition of the relative height
$h^\kappa(x)$ of a one-dimensional fluctuating interface indexed
by a continuously varying real paramater $0 \leq \kappa \leq 1$. It
interpolates between the height 
relative to the initial value ({\it i.e.} in $x=0$) when $\kappa = 0$
and the height relative to the spatially averaged height for $\kappa =
1$. We compute exactly the distribution $P^\kappa(h_m,L)$ of the
maximum $h_m$ of these relative heights for systems of finite size $L$
and periodic boundary conditions. One finds that it takes the scaling
form $P^\kappa(h_m,L) = L^{-1/2} f^\kappa (h_m L^{-1/2} )$ where the
scaling function $f^\kappa(x)$ interpolates between the Rayleigh
distribution for $\kappa=0$ and the Airy distribution for $\kappa=1$,
the latter being the probability distribution of the area under a
Brownian excursion over the unit interval. For arbitrary $\kappa$, one
finds that it is related to, albeit different from, the distribution
of the area restricted to the interval $[0, \kappa]$ under a Brownian
excursion over the unit interval.

\end{abstract}
\maketitle

\section{Introduction}


While the study of extreme value statistics (EVS) is by now a
longstanding issue in the fields of engineering \cite{gumbel}, finance
\cite{embrecht} or environmental sciences \cite{katz}, it was
recognized only recently that EVS plays a crucial role in the theory
of complex and disordered systems \cite{bouchaud_mezard}. Since then,
the understanding of EVS in spatially extended systems has been the
subject of many recent theoretical studies in statistical
physics. Indeed, although the EVS of a set of independent or weakly
correlated random variables is well understood, thanks to the
identification of a limited number of universality classes, much less
is known for the case of {\it strongly correlated} random
variables. In that respect, the study of fluctuating elastic
interfaces has recently attracted much attention \cite{GHPZ,
  RCPS,satya_mrh,satya_mrh2,GK1,GK2,Lee, schehr_sos, racz,
  burkhardt}. Although simpler to study, these models, in low
dimension, furnish an interesting example of strongly correlated
variables where analytical progress can be made.  
 
Here we are interested in the extremal statistics of the relative height of  a one-dimensional elastic line described, at equilibrium by the Gibbs-Boltzmann weight $P_{\rm st} \propto \exp{(-{\cal H})}$ with
\begin{eqnarray}\label{def_model}
{\cal H} = \frac{1}{2} \int_0^L \left( \frac{\partial H}{\partial
  x}\right)^2 dx \;, 
\end{eqnarray}
where $H(x)$ is the height of the interface and $L$ is the linear size
of the system. Such models (\ref{def_model}) have been extensively studied to describe various
experimental situations such as fluctuating step edges on crystals
with attachment/detachment dynamics of adatoms \cite{adatoms}. Here we
also impose periodic boundary conditions (pbc) such that $H(0) = H(L)$
and the process defined by ${\cal H}$ in Eq. (\ref{def_model}) is thus
a Brownian bridge, {\it i.e.} a Brownian motion constrained to start
and end at the same point. The Hamiltonian ${\cal H}$ is obviously
invariant under a global shift $H(x) \to H(x) + c$ and therefore a
more physically relevant quantity is the {\it relative} height. Here
{\it relative} means measured with respect to some (arbitrary)
reference value. Up to now two different reference values have been
considered : (i) the spatial average value
as in \cite{RCPS,satya_mrh,satya_mrh2,GK1, GK2,Lee, schehr_sos, racz} or
(ii) the height at $x=0$, $H(0)$ (or at any
other point in space because of pbc) which is a rather natural choice
in the context of time series \cite{burkhardt} (note that a 
recent work considered, for 
weakly correlated random variables, yet another choice where the reference is
the minimum value of the height field on $[0,L]$ \cite{moloney} which
we will not discuss here). This yields two distinct definitions of the
relative height 
\begin{eqnarray}\label{def_rel_init} 
(i) \;  h^1(x) = H(x) - L^{-1} \int_0^L H(x') \,dx' \;, \; (ii) \; \;
  h^{0}(x) = H(x) - H(0) \;.
\end{eqnarray} 
Once a reference value is chosen, one defines $h_m = \max_{0\leq x
  \leq L} h^{1}$ (or $h_m = \max_{0\leq x \leq L} h^{0}$) and we ask
the question : what is the probability density $f^{0,1}(h_m,L)$? It
is straightforward to see from Eq. (\ref{def_model}) that  
\begin{eqnarray}
&&(i) \; \langle  h^1(x_0) h^1(x_0+x) \rangle = \frac{L}{12} \left[1 -
    \frac{6x}{L}\left(1-\frac{x}{L}\right)   \right]  \;, \\
 &&(ii) \;  \langle
    h^0(x_0) h^0(x_0+x) \rangle = x_0 \left( 1- \frac{x+x_0}{L} \right) \;,
\end{eqnarray}
so that the variables $h^1(x)$ or $h^0(x)$ are obviously strongly
correlated : the computation of $h_m$ is thus non trivial. 

\begin{figure}[h]
\begin{center}
\includegraphics[width = 0.5 \linewidth, angle=-90]{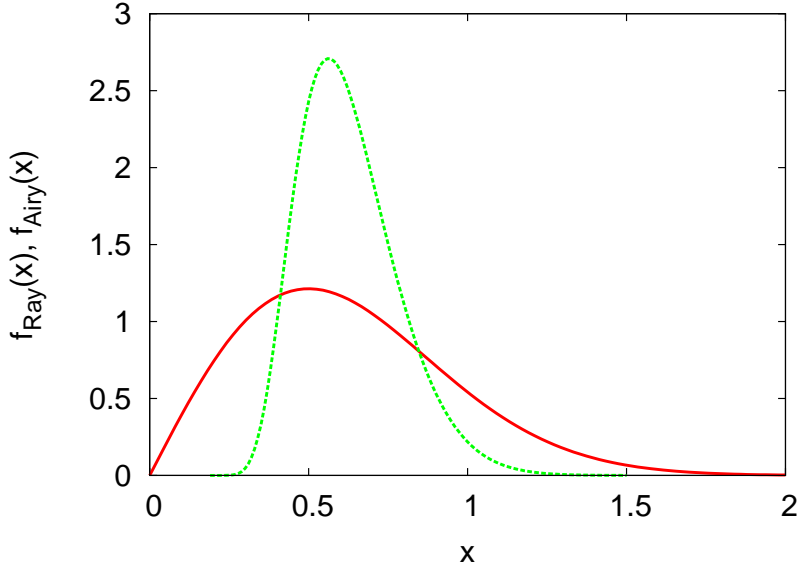}
\caption{The dashed green line represents the Airy distribution function $f^1=f_{\rm{Airy}}$:
this is the pdf of the maximum measured relatively to the spatial average value of heights of the interface on the interval $[0,1]$.
The solid red line represents the Rayleigh distribution function $f^0=f_{\rm{Ray}}$:
this is the pdf of the maximum measured relatively to the first point 
of the interface on the interval $[0,1]$.}
\label{plot_AiryRayleigh} 
\end{center}
\end{figure}

The distribution $f^{1}(h_m,L)$, studied numerically in
Ref.~\cite{RCPS}, was then computed analytically by Majumdar and
Comtet~\cite{satya_mrh, satya_mrh2} who showed, for the model described in Eq.~(\ref{def_model}) with pbc,
that $P^1(h_m,L) = L^{-1/2} f_{\rm Airy}(h_m L^{-1/2})$ and computed
exactly $ f_{\rm Airy}(x)$. Its Laplace transform is given by~\cite{satya_mrh}
\begin{equation}
\int_0^{\infty} f_{\rm Airy}(x) e^{-px} dx = p\sqrt{2\pi} \sum_{n=1}^{\infty}
e^{-\alpha_n p^{2/3}2^{-1/3}} \:,
\label{lt1}
\end{equation} 
where $\alpha_n$'s are the magnitudes of the zeros of the 
standard Airy function ${\rm Ai}(z)$ on the negative real
axis~\cite{abramowitz}. For example, $\alpha_1=2.3381\dots$,
$\alpha_2=4.0879\dots$,  
$\alpha_3=5.5205\dots$ etc ~\cite{abramowitz}. It is then possible to
invert the Laplace transform \cite{satya_mrh2, takacs_invert}
\begin{eqnarray}\label{def_airy}
f_{\rm Airy}(x) = \frac{2\sqrt{6}}{x^{10/3}}\sum_{n=1}^\infty e^{-b_n/x^2}
b_n^{2/3} U(-5/6,4/3,b_n/x^2) \;,
\label{Airy_dist}
\end{eqnarray}
where $b_n = 2\alpha_n^3/27$ and $U(a,b,z)$ is the confluent
hypergeometric function \cite{abramowitz}. This inversion is useful as
it can then be evaluated numerically, and plotted as in
Fig.~\ref{plot_AiryRayleigh}. In particular, for small $x$ it behaves
as \cite{satya_mrh2}  
\begin{eqnarray} \label{airy_smallx}
f_{\rm Airy}(x) &\sim & \frac{8}{81} \alpha_1^{9/2} x^{-5}\, e^{- 2\alpha_1^3/{27 x^2}} \quad {\rm
  as}\quad x\to 0  \;,
\end{eqnarray}
while at large $x$, $f_{\rm Airy}(x) \sim (72 \sqrt{6/\pi}) x^2 e^{-6
  x^2}$ \cite{janson_louchard}.   

Interestingly, the Airy distribution in Eq.~(\ref{def_airy}) also describes
the distribution of the area under a Brownian excursion over the unit
interval. We recall that a Brownian excursion is a
Brownian motion starting and ending at the same point and constrained to
stay positive in-between, on the unit interval \cite{satya_mrh2,
  janson_review}. Surprisingly, this 
distribution arises in various seemingly unrelated problems in computer
science and graph theory \cite{satya_review}.   

On the other hand, T. W. Burkhardt {\it et al.} focused on
$f^0(h_m,L)$ \cite{burkhardt} which is different from the Airy
distribution. Indeed, it is easy to see that in that case the
distribution of $h_m$ is just the distribution of the maximum of a
Brownian bridge on the interval $[0,L]$. Therefore $P^0(h_m,L) =
L^{-\frac{1}{2}} f_{\rm Ray}(h_m L^{-\frac{1}{2}})$ where
\begin{eqnarray}\label{def_ray}
f_{\rm Ray}(x) = 4 \, x \, \exp\left(-2 x^2\right) \;,
\end{eqnarray}
which is known in the literature as the Rayleigh distribution. In the
literature on random matrix theory, $f_{\rm Ray}(x)$ (which is plotted in 
Fig. \ref{plot_AiryRayleigh}) is also known 
under the name of ``Wigner surmise'' as it describes the distribution
of the spacing 
between consecutive energy levels in the Gaussian Orthogonal Ensemble
(GOE). In Ref. \cite{burkhardt} it was also pointed out that the distribution of $h_m = 
\max_{0 \leq x \leq L}h(x)$, for generic translationnaly invariant
systems, is related to the distribution of near extreme events, which
has recently attracted some attention in statistical physics
\cite{sanjib_nearextreme}.  

Although the definition of these two relative heights are a priori
only slightly different (see Eq. (\ref{def_rel_init})), the
distribution of their maximum in Eq. (\ref{def_airy}) and in
Eq. (\ref{def_ray}) has 
completely different expressions. These two definitions of the maximal
relative heights thus lead to qualitatively different distributions :
this is particularly striking if one considers their small argument behavior in
Eq. (\ref{airy_smallx}) and Eq. (\ref{def_ray}). The goal of this
paper is to understand how 
the distribution of the maximum evolves when one changes the
definition of the relative height from $h^1(x)$ to $h^0(x)$. In that
purpose, we choose an alternative reference value, indexed by a
continuously varying parameter $0 \leq \kappa \leq 1$ to define the
relative height $h^\kappa(x)$ as
\begin{eqnarray}\label{def_relkappa}
h^\kappa(x) = H(x) - \frac{1}{\kappa L}\int_0^{\kappa L} H(x') \, dx' \;
\;,
\end{eqnarray}
such that $\lim_{\kappa \to 1} h^\kappa(x) = h^{1}(x)$ and
$\lim_{\kappa \to 0} h^\kappa(x) = h^{0}(x)$ : $h^\kappa(x)$ thus
interpolates between $h^1(x)$ and $h^0(x)$. Using path integral
methods, we compute exactly the distribution $P^\kappa(h_m,L)$ of the
maximal relative height $h_m = \max_{0 \leq x \leq L} h^\kappa(x)$ and
find that $P^\kappa(h_m,L) = L^{-\frac{1}{2}} f^\kappa(h_m
L^{-\frac{1}{2}})$ where $ f^\kappa(x)$, which interpolates
between the Airy distribution for $\kappa = 1$ and the Rayleigh
distribution for $\kappa = 0$ is given by
\begin{eqnarray}\label{res_fkappa}
f^\kappa(x) = \frac{4 \sqrt{6}}{\pi} \, x^{-10/3} &&\int_0^{\infty} \rmd q \sum_{n=1}^{\infty} \left[ F_n(q) \right]^2 \
b_n(q,\kappa)^{2/3} \, \rme^{-b_n(q,\kappa)/x^2} \, \nonumber \\
&&\times U(-5/6,4/3,b_n(q,\kappa)/x^2) \;, 
\end{eqnarray}
where $b_n(q, \kappa) = (2\kappa/27)(\alpha_n +
(1-\kappa)q^2/\kappa)^3$ and where the coefficients $F_n(q)$ are given
below (\ref{Fnk}). In particular, for small argument, we show that it
behaves as
\begin{eqnarray}
f^\kappa(x) \sim \left(\frac{\kappa}{1-\kappa}\right)^{3/2}
x^{-2} \exp\left(-\frac{2\alpha_1^3}{27} \frac{\kappa}{x^2} \right) \;.
\end{eqnarray}    

The paper is organized as follows. In section II, we illustrate the
method of path integral to compute the distribution $f^0(x)$, {\it
  i.e.} the distribution of the maximum of a Brownian bridge. In
section III, we use this path-integral formalism to compute the
distribution of the maximum of the relative heights $h^\kappa(x)$
defined in Eq. (\ref{def_relkappa}). In section IV, we analyse in
detail the limits $\kappa \to 1$ and $\kappa \to 0$ of the scaling
function $f^\kappa(x)$ in Eq. (\ref{res_fkappa}) and in section
V, we present the comparison of our exact results with numerical
simulations for different values of $0 \leq \kappa \leq 1$. In section
VI, guided by the fact that $f^1(x) =  f_{\rm Airy}(x)$ describes
the distribution of the 
area under a Brownian excursion over a unit interval, we show that
$f^\kappa(x)$ is related to, albeit different from, the distribution of the area restricted to the interval $[0, \kappa]$
under a Brownian excursion over the unit interval. Finally, we
conclude in Section VII. For clarity, some technical details have been left in
Appendix.

\section{Distribution of the maximum of a brownian bridge : a
  path-integral approach}
\label{max_bridge}

In this section, we derive the distribution of the maximal relative height $h_m = \max_{0 \leq x \leq L } h^{0}(x)$ where $h^0(x) = H(x) - H(0)$. The joint distribution of these heights $h^0(x)$, given Eq. (\ref{def_model}), is
\begin{eqnarray} 
\label{poids}
P[\{h^0\}]	= Z_L^{-1} \exp \left(-\half \int_{0}^{L} \left(\frac{\partial h^0(x)}{\partial x}\right)^2 \rmd x \right) \delta(h^0(0)-h^0(L)) \delta(h^0(0)) \;, \nonumber \\
\end{eqnarray}
where $Z_L$ is a normalization constant. In this expression (\ref{poids}), the first delta function ensures the pbc while the second one ensures simply the definition of $h^0(x)$ in Eq. (\ref{def_rel_init}). From this statistical weight, one directly obtains the cumulative distribution of the maximum $F^0(h_m,L) = {\rm Prob}{[\max_{0\leq x\leq L}{h^0(x)} < h_m,L]}$ as
\begin{equation}
\label{wigner_cumul}
F^0(h_m,L)=  \frac{1}{Z_L} \int\limits^{h^0(L)=0}_{h^0(0)=0}
 {\mathcal D}h(x)  e^{-\half \int_{0}^{L} (\partial_x h^0(x))^2}
 \times \prod_{0 \leq x \leq L} \theta (h_m-h^0(x)) \;,
\end{equation}
where product of Heaviside step functions constraint the paths to stay under $h_m$. Note that the normalization constant $Z_L$ has just the same path-integral representation as in Eq. (\ref{wigner_cumul}) but without the constraint ensured by the step-functions 
\begin{eqnarray}\label{normalization_bb}
Z_L = \int\limits^{h^0(L)=0}_{h^0(0)=0}
 {\mathcal D}h(x)  e^{-\half \int_{0}^{L} (\partial_x h^0(x))^2} \;.
\end{eqnarray}
One sees explicitly on these expressions (\ref{wigner_cumul}, \ref{normalization_bb}) that $F^0(h_m,L)$ is the cumulative distribution of the maximum of a Brownian bridge on the interval $[0,L]$. One can now use path-integrals techniques to write $F^0(h_m,L)$ as
\begin{equation}
\label{cumul_pont_propag}
F^0(h_m,L)=\frac{ \langle 0 |  \exp(- L \hat H_m) \, | 0 \rangle }{
  \langle 0 | \, \exp(-L \hat H_{\rm{free}}) \,| 0 \rangle } ,
\end{equation}
where $\hat H_{\rm{free}}= -\half \frac{d^2}{dx^2}$ and $\hat H_m
= - \half \frac{d^2}{dx^2} + V_m(x)$ where $V_m(x)$ is a confining
potential such that $V_m(x) = 0$ if $-\infty < x < h_m$ and $V_m(x) =
+ \infty$ if $x > h_m$. The numerator and the denominator in
Eq. (\ref{cumul_pont_propag}) can be computed in a similar way: if
one denotes $| E \rangle$ the eigenvectors of a Hamiltonian $\hat H$,
with eigenvalues $E$, $\hat H |E\rangle = E |E \rangle$ then the
propagator can be expanded on the eigenbasis as
\begin{eqnarray}\label{propag_eigen}
\langle y | \exp(-L \hat H) | x \rangle = \sum_{E} \psi_E^*(y) \, \psi_E(x)
\ \rme^{-L E} \; ,
\end{eqnarray}
where $\psi_E(x) = \langle x | E \rangle$. Applying this formula
(\ref{propag_eigen}) to $\hat H_m$ one has
\begin{eqnarray}
\langle 0 |  \exp(- L \hat H_m) \, | 0 \rangle = \frac{2}{\pi}
\int_0^\infty dq (\sin (q \, h_m))^2 e^{-\frac{q^2}{2}} =
\frac{1}{\sqrt{2 L \pi}}\left(1- e^{-2 \frac{h_m^2}{2L}}\right) \;,
\end{eqnarray}
and similarly $\langle 0 |  \exp(- L \hat H_{\rm free}) \, | 0 \rangle
= (2 \pi L)^{-\half}$. And finally one obtains the cumulative
distribution function as
\begin{eqnarray}
F^0(h_m,L) = 1- e^{-2 \frac{h_m^2}{2L}} \;,
\end{eqnarray}
from which one deduces that the probability distribution function
$P^0(h_m,L) = \partial_{h_m} F^0(h_m,L)$ takes the scaling form
(valid for all $L$ for this model) 
\begin{eqnarray}\label{def_rayleigh}
P^0(h_m,L) = \frac{1}{\sqrt{L}} f^0 \left(\frac{h_m}{\sqrt{L}} \right) \;, \;
f^0(x) = 4 x e^{-2x^2} \;,
\end{eqnarray}
which is known in the literature as the Rayleigh function. Of course
there are other simpler methods to compute $f^0(h_m,L)$ for the
present case (see for instance~\cite{burkhardt}). However the path integral method presented here can be extended to
more complicated situations, and will be used below.

\section{Distribution of the Maximum Relative Height $P^\kappa(h_m,L)$}
\label{h_kappa}

In this section, we extend the calculation presented above to the
computation of the cumulative distribution of the maximal relative
height $h_m = \max_{0 \leq x \leq L } h^{\kappa}(x)$ where
$h^\kappa(x)$ is defined in Eq. (\ref{def_relkappa}). As above we
start with the joint distribution of the heights $h^\kappa(x)$, which
from Eq. (\ref{def_model}), can be computed as
\begin{eqnarray} 
\label{poids_kappa}
P[\{h^\kappa\}]	=&& \frac{1}{\tilde Z_{\kappa}} \exp \left(-\half \int_{0}^{L}
\left(\frac{\partial h^\kappa(x)}{\partial x}\right)^2 \rmd x \right)
\delta(h^\kappa(0)-h^\kappa(L)) \nonumber \\
&& \times \delta \left( \int_0^{\kappa L} h^\kappa(x') dx' \right) \;,
\end{eqnarray}
where $\tilde Z_\kappa$ is a normalization constant which ensures that $\int
\mathcal{D}h^\kappa(x) P[\{h^\kappa\}] = 1$. The cumulative
distribution of the maximum $F^\kappa(h_m,L)$ can thus be written as
\begin{eqnarray}\label{starting_kappa}
F^\kappa(h_m,L)=  && \frac{1}{\tilde Z_{\kappa}} \int_{-\infty}^{h_m} \rmd u
 \int\limits^{h^\kappa(L)=u}_{h^\kappa(0)=u} 
 {\mathcal D}h^\kappa(x)  e^{-\half \int_{0}^{L} (\partial_x h^\kappa(x))^2 \rmd x} \nonumber \\
 && \times \prod_{0\leq x \leq L} \theta (h_m-h^\kappa(x)) \
 \delta\left(\int_0^{\kappa L} h^\kappa(x') dx' \right)\;,
\end{eqnarray}
which is the starting point of our exact computation.

\subsection{Normalization constant}
\label{normalization_constant}

As above (see section~\ref{max_bridge}), the normalization constant $\tilde Z_{\kappa}$ 
has the same path integral expression as Eq.~(\ref{starting_kappa}), but without
the constraint that each path must stay below $h_m$, that is
\begin{equation}
\tilde Z_{\kappa}=\int_{-\infty}^{+\infty} \rmd u \int\limits^{h^\kappa(L)=u}_{h^\kappa(0)=u} 
 {\mathcal D}h^\kappa(x)  e^{-\half \int_{0}^{L} (\partial_x h^\kappa(x))^2 \rmd x}
 \delta\left(\int_0^{\kappa L} h^\kappa(x') dx' \right)\; .
\end{equation}
Let us make the change of variable $y(x)=h^\kappa(x)-u$ to obtain
\begin{equation}
\tilde Z_{\kappa}=\int_{-\infty}^{+\infty} \rmd u \int\limits^{y(L)=0}_{y(0)=0} 
 {\mathcal D}y(x)  e^{-\half \int_{0}^{L} (\partial_x y(x))^2 \rmd x}
 \delta\left(\int_0^{\kappa L} y(x') dx' -\kappa L u \right)\; .
\end{equation}
The path integral is, up to a normalization constant, 
the probability that the truncated area $A_{\kappa}=\int_0^{\kappa L} y(x) \rmd x$
under a Brownian bridge $\{y(x),\ 0\leq x \leq L\}$ is equal to $\kappa
L u$. One thus has
\begin{equation}\label{def_proba}
{\rm Proba}(A_{\kappa}=\kappa L u, L)= \frac{\int\limits^{y(L)=0}_{y(0)=0} 
 {\mathcal D}y(x)  e^{-\half \int_{0}^{L} (\partial_x y(x))^2 dx}
 \delta\left(\int_0^{\kappa L} y(x') dx' -\kappa L u \right)}{
\int\limits^{y(L)=0}_{y(0)=0} 
 {\mathcal D}y(x)  e^{-\half \int_{0}^{L} (\partial_x y(x))^2 \rmd x}}\; .
\end{equation}
The denominator is just $(2 \pi L)^{-1/2}$, as computed before (in Section~\ref{max_bridge}). 
Then we simply have
\begin{equation}
\tilde Z_{\kappa}=\frac{1}{\sqrt{2\pi L}} \int_{-\infty}^{+\infty}
\rmd u \,{\rm Proba}(A_{\kappa}=\kappa L u,L) \nn
\end{equation}
Changing $v=\kappa L u$,  and using the property that $P(A_{\kappa},L)$
is, by definition, normalized (\ref{def_proba}), we find that
\begin{equation}
\label{zkappa}
\tilde Z_{\kappa}=\frac{1}{\sqrt{2\pi L} \kappa L} .
\end{equation}

\subsection{Cumulative distribution of the maximal relative height (MRH)}

Proceeding as in the first section, we now compute the cumulative
distribution of the MRH using path integrals. We start with the
expression given in Eq. (\ref{starting_kappa})   
\begin{eqnarray}
\label{expr_cumul}
F^\kappa(h_m,L)=  && \frac{1}{\tilde Z_{\kappa}} \int_{-\infty}^{h_m} du
 \int\limits^{h^\kappa(L)=u}_{h^\kappa(0)=u}
 {\mathcal D}h^\kappa(x)  e^{-\half \int_{0}^{L} (\partial_x h^\kappa(x))^2 \rmd x} \nonumber \\
 &&\times \prod_{0 \leq x \leq L} \theta (h_m-h^\kappa(x)) \delta \left( \int_0^{\kappa L} dx' h^\kappa(x') \right)\;,
\end{eqnarray}
where $\tilde Z_\kappa$ is given above~(\ref{zkappa}). The first
integral in Eq.~(\ref{expr_cumul}) indicates that we sum over all
initial points $u$ (which has to be below $h_m$), and then we
sum over all paths beginning in $u$ and constrained to stay entirely
under the height $h_m$ (imposed by the product of step
functions). Performing the change of variable $y(x)=h_m-h^\kappa(x)$
and $v=h_m-u$, we get all $h_m-$dependence in the delta function, and
we constraint the paths $y(x)$ to be positive: 
\begin{eqnarray}
F^\kappa(h_m,L) = && \kappa L\sqrt{2 \pi L}\,  \  \int\limits_{0}^{+\infty} \rmd v \int\limits_{y(0)=v}^{y(L)=v} \!\!\!\!\! \mathcal{D}y(x)  \rme^{-\half \int_0^L \rmd x (\partial_{x} y)^2} \nonumber \\
&& \times \delta\left( h_m \kappa L - \int_0^{\kappa L} y(x') \rmd x' \right) \ \prod_{0\leq x\leq L} \theta(y(x) ) \;.
\end{eqnarray}
The constraint of positivity can be incorporated rather easily by adding a potential term $V_0(y)$ in the exponential, with $V_0(y)=+\infty$ if $y<0$ and $V_0(y)=0$ if $y>0$. Then we have
\begin{eqnarray}
 F^\kappa(h_m,L) = && \kappa L\sqrt{2 \pi L}\,  \ \int\limits_{0}^{+\infty} \rmd v \int\limits_{y(0)=v}^{y(L)=v} \!\!\!\!\! \mathcal{D}y(x) \
 \rme^{-\int_0^T \rmd x \left\{ \half (\partial_{x} y)^2 +V_0(y(x)) \right\} } \nonumber \\
&& \times \delta\left( \kappa L h_m - \int_0^{\kappa L} y(x) \rmd x \right)  \;.
\end{eqnarray}
Using $\delta(\alpha x) = (1/\alpha) \delta (x)$, we isolate the variable $h_m$, and we write
\begin{eqnarray}
F^\kappa(h_m,L) = && \sqrt{2 \pi L}\  \int\limits_{0}^{+\infty} \rmd v \int\limits_{y(0)=v}^{y(L)=v} \!\!\!\!\! \mathcal{D}y(x) 
 e^{-\int_0^L \rmd x \left\{ \half (\partial_{x} y)^2 +V_0(y(x)) \right\} }  \nonumber \\
&&\times \delta\left( h_m  - \frac{1}{\kappa L} \int_0^{\kappa L} y(x') \rmd x'
\right)  \;,
\end{eqnarray}
and all the dependence on $h_m$ is contained in the
$\delta$-function. Thus its Laplace transform with respect to $h_m$,
$\tilde{F}^\kappa(p,L)=\int_0^\infty \rmd h_m F^\kappa(h_m,L) \rme^{-p
  h_m}$ (notice that $h_m$ is necessarily positive) has a simple
expression  
\begin{eqnarray}
\tilde{F}^\kappa(p,L) = \sqrt{2 \pi L}\  \int\limits_{0}^{+\infty} \rmd v  \int\limits_{y(0)=v}^{y(L)=v} \!\!\!\!\! \mathcal{D}y(x) 
e^{-\int_0^L \rmd x \left\{ \half (\partial_{x} x)^2 +V_0(y(x))
  \right\} } e^{- \frac{p}{\kappa L} \int_0^{\kappa L} y(x) \rmd x } \;. \nonumber \\
\end{eqnarray}
Let us call $u=y(\kappa L)$ the intermediary point and using the Markov property, we can separate the path integral in two independent blocks: one for the time interval $[0,\kappa L]$ and the other for the time interval $[\kappa L,L]$. We have
\begin{eqnarray}
\tilde{F}^\kappa(p,L) = \sqrt{2 \pi L}\  \int\limits_{0}^{+\infty} \rmd v \int\limits_{0}^{+\infty} \rmd u 
\int\limits_{y(0)=v}^{y(\kappa L)=u} \!\!\!\!\! \mathcal{D}y(x) 
e^{- \int_0^{\kappa L} \rmd x \left\{ \half (\partial_{x} y)^2 +V_0(y(x)) + \frac{p}{\kappa L}  y(x) \right\} } \nonumber \\
\times \int\limits_{y(\kappa L)=u}^{y(L)=v} \!\!\!\!\! \mathcal{D}y(x)
\  e^{-\int_{\kappa L}^L \rmd x \left\{ \half (\partial_{x} y)^2
  +V_0(y(x)) \right\} }  \;. 
\end{eqnarray}
The first factor is the propagator  between time $0$ and $\kappa L$ of
a quantum particle in the triangular potential $V_{0,p/(\kappa
  L)}(x)=V_0(x)+\frac{p}{\kappa L} x$, denoted as
$G_{\rm{Airy}}(u,\kappa L|v,0)$ (the index `Airy' refers to the
fact that the solution of the Schr\"odinger equation with a triangular
potential $V_{0,p/(\kappa L)}(x)$ is solved by Airy functions), and
the second term, denoted as  $G_{0}(v,L|u,\kappa L)$ is simply the
propagator of a quantum particle in the potential $V_0(x)$ between the
times $\kappa L$ and $L$. One thus has 
\begin{equation}
\label{cumul_propagateurs}
\tilde{F}^\kappa(p,L) = \sqrt{2 \pi L}\  \int\limits_{0}^{+\infty} \rmd v \int\limits_{0}^{+\infty} \rmd u 
\left\{ 
G_{0}(v,L|u,\kappa L) \times G_{\rm{Airy}}(u,\kappa L|v,0)
\right\} \;,
\end{equation}
where the $p-$dependence is entirely contained in the Airy propagator. Note that our calculation is completely justified when $0 < \kappa < 1$, because of the cut in two parts of the path integral. We will examine in more detail the two limiting cases $\kappa \to 0$ and $\kappa \to 1$ below. 

The probability density function of the MRH $P^\kappa(h_m,L)$ is
simply the derivative of $F^\kappa(h_m,L)$ with respect to $h_m$. Its Laplace transform 
$\tilde{P}^\kappa(p,L) = \int_0^\infty e^{-p h_m} P^\kappa(h_m,L) \, dh_m$ is thus $\tilde{P}^\kappa(p,L)=p \tilde{F}^\kappa(p,L)$,
\begin{equation}
\tilde{P}^\kappa(p,L) = p\sqrt{2 \pi L} \  \int\limits_{0}^{+\infty} \rmd v \int\limits_{0}^{+\infty} \rmd u 
\left\{ 
G_{0}(v,L|u,\kappa L) \times G_{\rm{Airy}}(u,\kappa L|v,0)
\right\}.
\end{equation}
We first start with $G_{0}(v,L|u,\kappa L)$ which is the propagator of a particle with a wall at the origin. It is given by
\begin{eqnarray}
\label{propagateur_02}
G_0(v,L|u,\kappa L)  &=& \frac{1}{\sqrt{2 \pi L(1-\kappa)}} \left( \rme^{- \frac{(u-v)^2}{2 L(1-\kappa)}} - e^{-\frac{(u+v)^2}{2 L(1-\kappa)}} \right) \\
&=& \frac{2}{\pi} \int_0^{+\infty} \rmd q \, \sin(qv) \sin(qu) \, \rme^{-\frac{L(1-\kappa)}{2} q^2} \;,
\end{eqnarray}
where the expression in the second line will be useful in the following. The propagator $G_{\rm{Airy}}(u,\kappa L|v,0)$ can  be computed using the formula in Eq. (\ref{propag_eigen})(see also in \ref{app_prop_airy}, formula~(\ref{airy_formule1})):
\begin{eqnarray}
\label{propagateur_airy}
&& \nonumber \\
&&\hspace*{-1.5cm}G_{\rm{Airy}}(u,\kappa L|v,0) = \left( \frac{2p}{\kappa L}\right)^{1/3}\sum_{n=1}^{\infty} 
\frac{\ind{Ai}\left( \left( \frac{2p}{\kappa L}\right)^{1/3} u - \alpha_n \right) \ind{Ai}\left( \left( \frac{2p}{\kappa L}\right)^{1/3} v - \alpha_n \right) }
{\left(\ind{Ai}'[-\alpha_n] \right)^2} \nonumber \\
&& \times e^{- \alpha_n 2^{-1/3} (p\sqrt{\kappa L})^{2/3} } \;.
\end{eqnarray}
We then perform the changes of variables $v \to (2p/(\kappa L))^{-1/3} v$ and $u \to (2p/(\kappa L))^{-1/3} u$ followed by $q \rightarrow (2p/(\kappa L))^{1/3} q$
to obtain
\begin{eqnarray}
&&\hspace*{-2.4cm}\tilde{P}^\kappa(p,L) = p \sqrt{\frac{8L}{\pi}} \int\limits_0^\infty \rmd u \int\limits_0^\infty \rmd v \int\limits_0^\infty \rmd q  
\sin (qu) \sin (qv) \exp \left[- 2^{-1/3} (1-\kappa) \kappa^{-2/3} (p\sqrt{L})^{2/3} q^2 \right] \nonumber \\
&&\times \sum_{n=1}^{\infty} \frac{ \ind{Ai}(u-\alpha_n) \ind{Ai}(v-\alpha_n) }{[\ind{Ai}'(-\alpha_n)]^2 } \exp\left[-\alpha_n 2^{-1/3} \kappa^{1/3} (p\sqrt{L})^{2/3} \right] \;.
\end{eqnarray}
One recognizes the scaling variable $p \sqrt{L}$ in the Laplace space which implies
that the scaling variable in the real space is, as expected,
$h_m/\sqrt{L}$, and then we can write the scaling law 
\begin{equation}\label{scaling_law}
P^\kappa(h_m,L)=\frac{1}{\sqrt{L}} {f}^\kappa\left( \frac{h_m}{\sqrt{L}} \right) ,
\end{equation}
where the Laplace transform of $f^{\kappa}(x)$ with respect to $x$ is given by
\begin{eqnarray}
\label{scaling_laplace1}
&&\hspace*{-2.4cm}\tilde{f}^\kappa(p) = p \,\sqrt{\frac{8}{\pi}} \int\limits_0^{\infty} \rmd q \int\limits_0^{\infty} \rmd u \int\limits_0^{\infty} \rmd v  \
\sin(qu) \sin(qv) \ \exp\left[ - \half (1-\kappa) \left( \frac{\kappa}{2} \right)^{-2/3} p^{2/3} q^2 \right]  \nonumber \\
&&\times \sum_{n=1}^{\infty}  \frac{\ind{Ai}\left( u - \alpha_n \right) \ind{Ai}\left( v - \alpha_n \right) }
{\left(\ind{Ai}'[-\alpha_n] \right)^2} \
\exp\left[- \alpha_n 2^{-1/3} \kappa^{1/3} p ^{2/3} \right] .
\end{eqnarray}
This formula can be written in the more compact form
\begin{equation}
\label{interpol_echelle}
\tilde{f}^\kappa(p) = \sqrt{\frac{8}{\pi}} \int\limits_0^{\infty} \rmd q
\sum_{n=1}^{\infty} \left[F_n(q)\right]^2 
p\, \rme^{-2^{-1/3} p^{2/3} \left( \alpha_n \kappa^{1/3} + (1-\kappa)
  \kappa^{-2/3} q^2 \right) }, 
\end{equation} 
with 
\begin{equation}\label{Fnk}
F_n(q)=\int\limits_0^{\infty} \rmd u \frac{\ind{Ai}(u-\alpha_n)}{\ind{Ai}'(-\alpha_n)} \sin(qu) .
\end{equation}
To invert the Laplace transform in Eq. (\ref{interpol_echelle}) we
will make use of the result found by Takacs in
Ref. \cite{takacs_invert} 
\begin{equation}
\mathcal{L}^{-1} \left[ \sqrt{2 \pi} \, p \, \rme^{-2^{-1/3} p^{2/3} \gamma_n} \right] (x) =
\frac{2\sqrt{6}}{x^{10/3}} \, \rme^{-b_n / x^2} b_n^{2/3} U(-5/6,4/3,b_n/x^2) ,
\end{equation}
where $b_n=2 \gamma_n^3 /27$ and $U(a,b,z)$ is the confluent
hypergeometric function \cite{abramowitz}. This yields here 
\begin{eqnarray}
\label{distrib_real}
f^\kappa(x) = && \frac{4 \sqrt{6}}{\pi} \, x^{-10/3} \int_0^{\infty} \rmd q \sum_{n=1}^{\infty} \left[ F_n(q) \right]^2 \
b_n(q,\kappa)^{2/3} \, e^{-b_n(q,\kappa)/x^2} \nonumber \\ 
&& \times U(-5/6,4/3,b_n(q,\kappa)/x^2) .
\end{eqnarray}  
where $b_n(q,\kappa) = \frac{2 \kappa}{27} \left( \alpha_n +
\frac{1-\kappa}{\kappa} q^2 \right)^3$. This class of distributions
indexed by the real parameter $0 \leq \kappa \leq 1$ smoothly
interpolates between the Rayleigh distribution for $\kappa \to 0$ and
the Airy distribution when $\kappa \to 1$. This will be demonstrated
analytically (in particular the behavior when $\kappa \to 0$ is not
obvious on Eq. (\ref{distrib_real})) in the next section and numerically in
section VI.

\section{Recovering Rayleigh and Airy distributions}

In this section, we show how $f^\kappa(x)$ yield back -- as it should
-- the Rayleigh distribution (respectively the Airy distribution) in
the limit $\kappa \to 0$ (respectively in the limit $\kappa \to 1$).

\subsection{The limit $\kappa \to 0$ and the Rayleigh distribution}

We first study the limit $\kappa \to 0$. From the definition of
$h^\kappa(x)$ given in Eq. (\ref{def_relkappa}) one expects that it
yields back the definition of $h^0(x)$, {\it} i.e. the relative height
relative to $H(0)$ in Eq.~(\ref{def_rel_init}). This is also rather
obvious on the joint distribution in Eq.~(\ref{poids_kappa}), together
with $\tilde Z_\kappa \propto 1/\kappa L$, that it yields the joint
distribution of a Brownian Bridge. Therefore one expects indeed
that $f^\kappa(x)$ will converge to the Rayleigh distribution $f_{\rm
  Ray}(x)$ in Eq. (\ref{def_ray}). However, this limit $\kappa \to 0$
in our formulae for $f^\kappa(x)$, or in its Laplace transform 
$\tilde{f}^\kappa(p)$, is not trivial. This is partially due to the
fact that the computation of Rayleigh distribution has been performed
in ``direct'' space while for the Airy distribution, the computation
is more conveniently performed in Laplace variable. Indeed, one needs
an identity about the behavior of the Airy propagator when the time
difference tends to zero. In \ref{app_prop_lin}, we show the
identity (\ref{identity_airy}) which we then use straightforwardly in Eq.~(\ref{scaling_laplace1})
to compute the limit $\kappa \to 0$. 

Indeed, from the Laplace transform of the scaling function in
Eq.~(\ref{scaling_laplace1}), one performs the change of variables 
$v'=(2 p / \kappa)^{-1/3} v$, $u'=(2 p / \kappa)^{-1/3} u$, and $q'=(2 
p / \kappa)^{1/3}q$, so that we have 
\begin{eqnarray}
&&\hspace*{-2cm}\tilde{f}^{\kappa}(p)=p\sqrt{\frac{8}{\pi}} \int_0^\infty \rmd q' \int_0^\infty \rmd u' \int_0^\infty \rmd v'
\sin(q'u') \sin(q'v') \exp \left[-\half (1-\kappa) {q'}^2 \right]  \\
&&\times  \left(\frac{2p}{\kappa} \right)^{1/3} 
\sum_{n=1}^{\infty} \frac{ \textrm{Ai}\left[ \left( \frac{2p}{\kappa} \right)^{1/3} u'-\alpha_n \right] 
 \textrm{Ai}\left[ \left( \frac{2p}{\kappa} \right)^{1/3} v'-\alpha_n \right]}
 {\left( \textrm{Ai}'[-\alpha_n]\right)^2}
 e^{\alpha_n 2^{-1/3} \kappa^{1/3} p^{2/3}} . \nonumber
 \end{eqnarray}
 Now, the second line in that equation has a good limit when $\kappa \to
 0$, as shown in appendix~\ref{app_prop_lin}. We then make use of the
 identity in Eq.~(\ref{identity_airy}) to obtain the limit $\kappa \to
 0$ (with $u=u'$, $v=v'$, $q=q'$) 
 \begin{eqnarray}
 &&\hspace*{-2.4cm}\lim_{\kappa \to 0} \tilde{f}^{\kappa}(p) = 
 p\sqrt{\frac{8}{\pi}} \int_0^\infty \rmd q \int_0^\infty \rmd u \int_0^\infty \rmd v
 \sin(q u) \sin(q v) \exp \left[-\frac{{q}^2}{2} \right] 
 \times \delta (u-v) e^{-p u} \nonumber \\
 &&=  p\sqrt{\frac{8}{\pi}} \int_0^\infty \rmd u \ e^{- p u}  \int_0^\infty \rmd q \sin^2(q u) e^{-q^2/2} = \int_0^\infty \rmd u \ 4u e^{-2u^2} \ e^{-p u} \;.
 \end{eqnarray}
Now we recognize immediately the scaling function
in real space, since $\tilde{f}^{0}(p)=\int_0^\infty \rmd u f^0(u)
e^{-pu}$ such that:
\begin{equation}
\lim_{\kappa \to 0} f^{\kappa}(x)= f^0(x)=4 xe^{-2x^2} \;,
\end{equation}
as expected.

\subsection{The limit $\kappa \to 1$}

The analysis of the limit $\kappa \to 1$ is straightforward and can be
done directly on the expression of our scaling function $f^\kappa
(x)$ in Eq.~(\ref{distrib_real}). In the limit $\kappa \to 1$ one has 
\begin{equation}
\lim_{\kappa \to 1} b_n(q;\kappa)=b_n(q;1)=b_n,
\end{equation}
independently of $q$. Therefore the only dependence left in the
integrand of Eq. (\ref{distrib_real}) is contained in $F_n(q)$. It is
straightforward to obtain the useful relation
\begin{eqnarray}\label{fnk_kappa1}
\int\limits_0^\infty \rmd q \left[ F_n(q)\right]^2 
		&=& \int\limits_0^\infty \rmd q \int\limits_0^\infty
		\rmd u \int\limits_0^\infty \rmd v \
		\frac{\ind{Ai}(u-\alpha_n)}{\ind{Ai}'(-\alpha_n)}
		\sin(qu)\frac{\ind{Ai}(v-\alpha_n)}{\ind{Ai}'(-\alpha_n)}
		\sin(qv) \nonumber \\
&=&  \frac{\pi}{2} \int\limits_0^\infty \rmd u
		\left(
		\frac{\ind{Ai}(u-\alpha_n)}{\ind{Ai}'(-\alpha_n)}
		\right)^2 = \frac{\pi}{2} \; , 
\end{eqnarray}
%
where, we have used in the last line, the identity $\int_0^\infty \rmd
u (\ind{Ai}(u-\alpha_n))^2 = (\ind{Ai}'(-\alpha_n) )^2$. Using this
relation (\ref{fnk_kappa1}) in the expression of $f^\kappa(x)$ we get
\begin{equation}
\lim_{\kappa \to 1} f^{\kappa}(x)= f_{\rm Airy}(x) = 2 \sqrt{6} \ x^{-10/3} \sum_{n=1}^{\infty} b_n^{2/3} \, \rme^{-b_n/x^2} \, U(-5/6,4/3,b_n/x^2),
\end{equation}
which is the Airy distribution function.

\section{Some properties of the distribution $f^\kappa(x)$.}

In this section, we derive some properties of the distribution $f^\kappa(x)$.

\subsection{Asymptotic limit of the scaling function}

In this subsection, we study the asymptotic limit of the scaling
function when $x\ll 1$, which corresponds to $h_m \ll L$. 
One main difference between Rayleigh and Airy distributions is the
singularity of the latter for $x=0$.  
The Rayleigh distribution is perfectly analytic when $x\to 0$ ($\sim x$),
in contrast to the Airy distribution which exhibits an essential
singularity $ f_{\rm Airy} \sim x^{-5} e^{- 2\alpha_1^3/{27 x^2}}$,
see Eq. (\ref{airy_smallx}). Here we show that $f^\kappa(x)$ exhibits
an intermediate behavior with an essential singularity when $x \to 0$ (see
Eq. (\ref{asymptotic_0}) below).

\par
We start with the expression for $f^\kappa(x)$ given in Eq. (\ref{distrib_real}) which we recall here
\begin{eqnarray}
\label{distrib_real_bis}
&&f^\kappa(x) = \frac{4 \sqrt{6}}{\pi} \, x^{-10/3} \int_0^{\infty} \rmd q \sum_{n=1}^{\infty} \left[ F_n(q) \right]^2 \
b_n(q,\kappa)^{2/3} \, e^{-b_n(q,\kappa)/x^2} \nonumber \\
&&  U(-5/6,4/3,b_n(q,\kappa)/x^2) .
\end{eqnarray}  
that we want to analyse in the asymptotic limit $x \to 0$.  To this
purpose, we perform a change of variable $q = x \tilde q$ such that
$b_n(x \tilde q, \kappa)/x^2$ has  
the following small $x$ expansion
\begin{eqnarray}\label{expansion_bn}
\frac{b_n(x \tilde q, \kappa)}{x^2} = \frac{2 \kappa \alpha_n}{27 x^2} + \frac{6\alpha_n^2(1-\kappa) \tilde q^2}{27 \kappa}  + {\cal O}(x^2) \;,
\end{eqnarray}
so that, due to the exponential factor $\exp{(-b_n(x \tilde q;\kappa)/x^2)}$ in Eq. (\ref{distrib_real_bis}), only the term with $n=1$ can be retained in sum over $n$. Besides, using that $U(-5/6,4/3,y) \sim y^{5/6}$ for $y \gg 1$, 
\cite{abramowitz} we have 
\begin{equation}
\label{asymptotic_0}
f^\kappa(x) \approx \sqrt{\frac{2}{9 \pi} } {\alpha_1}^{3/2} \, [\chi(\alpha_1)]^2 \, \left(\frac{\kappa}{1-\kappa}\right)^{3/2} \, x^{-2} \, 
                        \exp\left[-\frac{2{\alpha_1}^3}{27} \, \frac{\kappa}{x^2} \right] \;,
\end{equation}
with $\chi(\alpha_1) = \int_0^\infty \rmd u u {\rm Ai}(u-\alpha_1)/{\rm
  Ai}'(-\alpha_1) = 1.8173...$ where we have used the small argument 
  behavior of $F_1(q)$ given in Eq. (\ref{Fnk}) $F_1(x \tilde q) =
  \chi(\alpha_1) (x \tilde q)+ {\cal O}(x^2)$. Of course, such formula
  is only valid for $0<\kappa<1$ strictly. One sees on this expression
  (\ref{asymptotic_0}) that this probability is exponentially small as
  soon as $h_m \ll  \sqrt{\kappa L}$. This behavior might be
  qualitatively understood by noticing that the configurations for
  which $h_m \ll \sqrt{\kappa L}$ are actually essentially {\it flat}
  on the interval $[0, \kappa L]$. 

A comparison between this asymptotic behavior in Eq. (\ref{asymptotic_0}) and the small argument behavior of the Rayleigh distribution (\ref{def_ray}) suggests that there exists a scaling regime $x \to 0$, $\kappa \to 0$ keeping the ratio $x/\sqrt{\kappa}$ fixed such that
\begin{equation}
f^\kappa(x) \propto x \ g\left( \frac{x}{\sqrt{\kappa}} \right),
\end{equation}
with the scaling function
\begin{equation}
g(u) \sim 
\cases{
{constant} & \textrm{when $u \to \infty$,} \\
u^{-3} \exp \left( - \frac{2 \alpha_1^3}{27} \frac{1}{u^2} \right) & \textrm{when $u \to 0$.}
}
\end{equation}
The limit $u\to \infty$ means that, for $x$ small but fixed, we take formally $\kappa \to 0$, and then we recover the asymptotic behavior of the Rayleigh distribution. On the other hand for $u \to 0$ we keep $x$ much smaller than $\kappa$ (actually $x \ll \sqrt{\kappa}$ as done previously), and one recovers the result above (\ref{asymptotic_0}). Similarly, a comparison between the behavior in Eq. (\ref{airy_smallx}) and Eq. (\ref{asymptotic_0}) suggests that there also exists a scaling regime where $x \to 0$ and $(1-\kappa) \to 0$ keeping $x/\sqrt{1-\kappa}$ fixed such that
\begin{equation}
f^\kappa(x) \propto x^{-5} \exp \left( - \frac{2 \alpha_1^3}{27} \frac{\kappa}{x^2} \right) \ h \left( \frac{x}{\sqrt{1-\kappa}} \right) \;,
\end{equation}
where the scaling function $h(v)$ has asymptotic behaviors
\begin{equation}
h(v) \sim
\cases{
\textrm{constant} & \textrm{when $v \to \infty$,} \\
v^3 & \textrm{when $v \to 0$.}
}
\end{equation}
Again $v\to \infty$ means that we keep $x$ fixed, but small, and we take $\kappa \to 1$, thus we recover the asymptotic behavior of the Airy distribution.
When $v\to 0$, we look at the region where $x \ll \sqrt{1-\kappa} \ll 1$, thus recovering the behavior obtained above (\ref{asymptotic_0}).

\subsection{Moments of the distribution.}

If one denotes the moments of $f^\kappa(x)$ as $M_n^\kappa = \int_0^\infty dx x^n f^\kappa(x)$, it is straightforward to show that, for the Rayleigh distribution, $M_n^{\kappa=0} = 2^{-n/2} \Gamma(1+n/2)$. On the other hand, the computation of the moments of the Airy distribution $M_n^{\kappa = 1}$ is highly non trivial. In Ref. \cite{takacs_invert}, Takacs found a recursive method to compute them and recently, M.J. Kearney {\it et al.} found an explicit integral representation of these moments as \cite{kearney}
\begin{eqnarray}
M_n^1 = \frac{4 \sqrt{\pi} n!}{\Gamma(\frac{3n-1}{2}) 2^{\frac{n}{2}}} \, K_n \;, \; K_n  = \frac{3}{4\pi^2} \int_0^\infty \frac{z^{\frac{3(n-1)}{2}}}{{\rm Ai}^2(z) + {\rm Bi}^2(z)} \, dz \;,
\end{eqnarray}
where ${\rm Ai}(z)$ and ${\rm Bi}(z)$ are the two linearly independent solutions of Airy's differential equation $y''(z) - z y(z) = 0$.

The computation of the moments $M_n^\kappa$ for arbitrary $\kappa$ seems to be highly complicated and we have been able to compute only the first one $M_1^\kappa = \int_0^\infty x f^{\kappa}(x) \, dx$. Indeed, permuting the integral over $x$ and the integral over $q$ together with the discrete sum over $n$ and  then performing the change of variable $x \to x \sqrt{b_n(q,\kappa)}$ one obtains that $M_1^\kappa$ is actually independent of $\kappa$, yielding
\begin{eqnarray}
M_1^\kappa = \frac{1}{2} \sqrt{\frac{\pi}{2}} \;, \; \forall \kappa \in [0,1] \;.
\end{eqnarray}

\section{Numerical results}

\begin{figure}[h]
\begin{minipage}{0.5\linewidth}
\includegraphics[angle=-90,width=\linewidth]{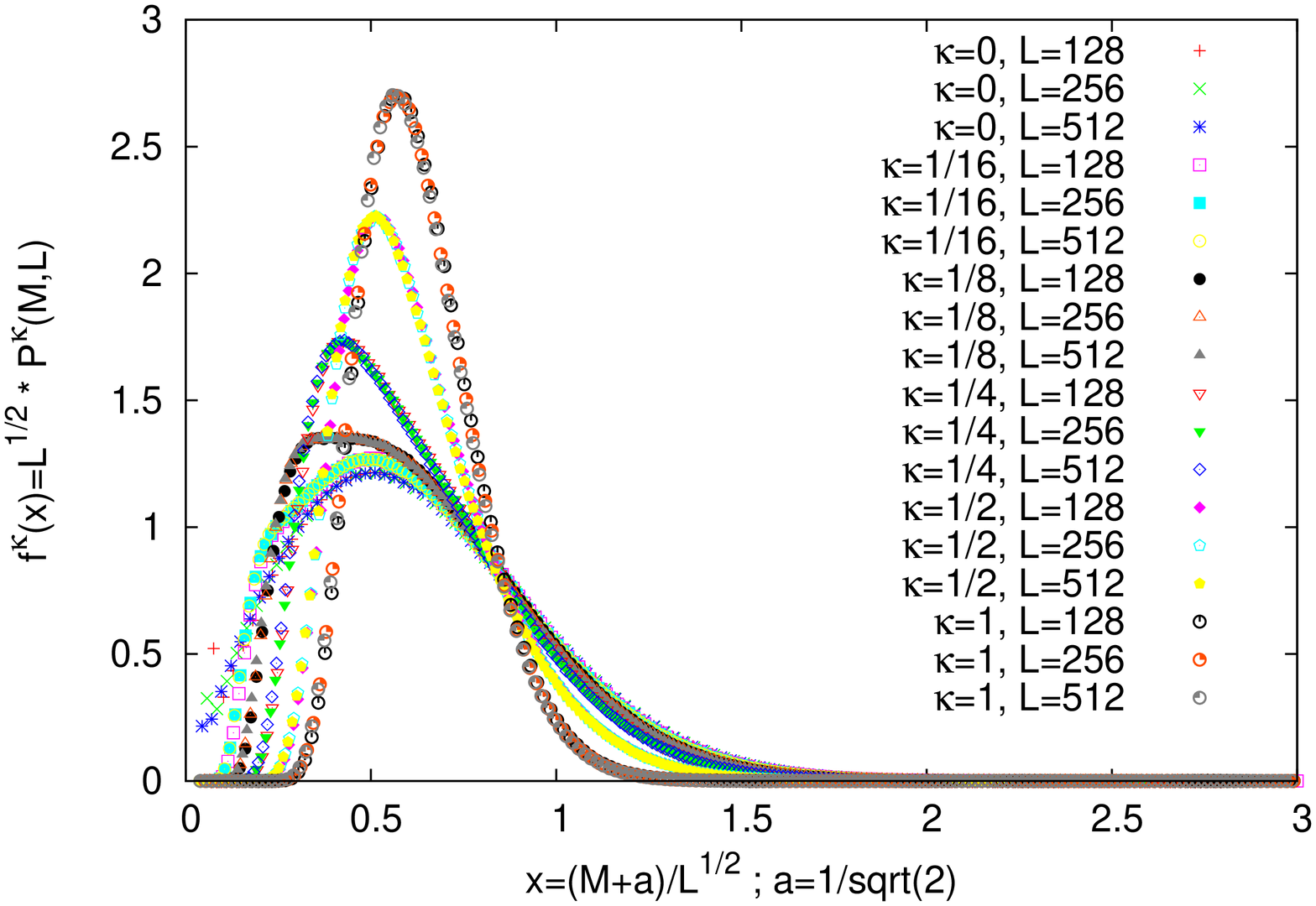}
\end{minipage}\hfill
\begin{minipage}{0.5\linewidth}
\includegraphics[angle=-90,width=\linewidth]{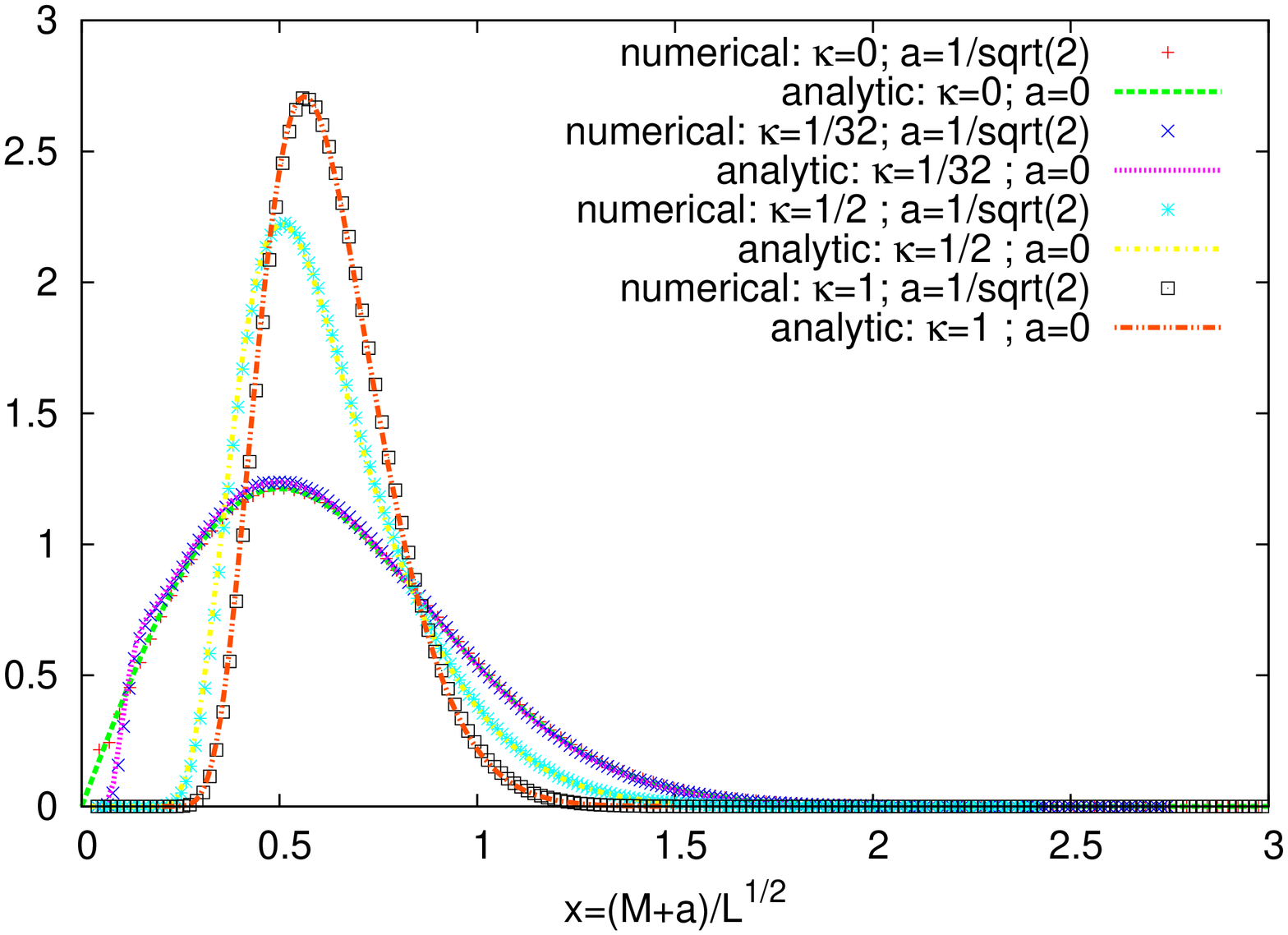}
\end{minipage}
\caption{{\bf Left :} Test of the scaling relation with numerical plots. From the raw
  histogram $P^\kappa_{\rm{num}}(h_m,L)$, we draw as co-ordinate
  $x=(h_m+a)/\sqrt{L}$, with $a=1/\sqrt{2}$  
and as ordinate $f^\kappa(x)=\sqrt{L} P_{\rm{num}}(h_m,L)$. 
Histograms with different $L$ but same $\kappa$ collapse in the same curve, which proves the scaling relation. {\bf Right} : Test of the analytic result. The dots are the numerical simulations (where we use the `translation' parameter $a$) 
and the curves represents numerical computations of the analytic function with the help of Mathematica.}\label{fig_scaling_comp}
\end{figure}

We have compared our analytical results with numerical simulations. To
generate the configurations of the interface distributed according to
the Gibbs-Boltzmann weight $P_{\rm st} \propto \exp{(-{\cal H})}$ with
${\cal H}$ given in Eq. (\ref{def_model}), one could of course use
standard Monte Carlo method. However, this suffers from critical
slowing down (with dynamical exponent $z=2$) and this is not very
efficient. Instead, as in Ref. \cite{satya_dasgupta}, one can use the
property that the process defined by this statistical weight
Eq. (\ref{def_model}) together with periodic boundary conditions is a
Brownian bridge. To generate it, one first generates numerically an
ordinary Brownian motion $B_0 = 0$ and $B_i = B_{i-1} + \eta_i$ where
$\eta_i$'s are independent and identically distributed (i.i.d.) random
variables drawn from a distribution ${\rm Proba}(\eta = x) = g(x)$. One can
then generate a Brownian Bridge through the relation $H_i = B_i -
(i/L) B_L$, which ensures pbc. One can show that this procedure
yields the correct statistical weight for large system size $L$ (when
$g(x)$ is a Gaussian, this procedure is exact for all $L$). The
relative height is simply $h^\kappa_i = H_i - \frac{1}{\kappa L}
\sum_{j=1}^{\kappa L}H_j$ and the distribution of $h_m = \max_{0 \leq i
  \leq L} h^\kappa_i$ is then computed for different system sizes $L =
128, 256, 512$ by averaging over $10^7$ samples.     
%
%
%
%
From the raw histograms representing
$P_{\rm{num}}^{\kappa}(h_m,L)$, for a given $L$, we want to
extract 
the scaling function $f_{\rm{num}}^{\kappa}(x)$ to first check
that the data for different sizes $L$, with $\kappa$ fixed, satisfy
the scaling form that we have obtained analytically in
Eq.~(\ref{scaling_law}). If we plot directly $\sqrt{L}
P_{\rm{num}}^{\kappa}(h_m,L)$ and as a function of $x=h_m/\sqrt{L}$, 
one can observe some finite size corrections to this scaling
form. 

These finite size effects can be understood using the analysis
done in Ref.~\cite{schehr_sos} where the first finite size corrections to
the Airy distribution were computed (see also the discussion in Ref. \cite{racz_fs}). There it was found that
\begin{equation}
P^1_{\rm{discrete}}(h_m,L) = \frac{1}{\sqrt{L}} \left( f^1 \left( \frac{h_m}{\sqrt{L}} \right)
						+ \frac{a}{\sqrt{L}} {f^1}'  \left(\frac{h_m}{\sqrt{L}} \right) 
						+ \mathcal{O}\left(
						L^{-1}\right) \right)
\end{equation}
where $a$ is the non-universal constant (that depends on the details of the Hamiltonian), and $f^1=f_{\rm{Airy}}$ the Airy
distribution function. This first correction can be resummed as
\begin{equation}\label{fss}
P^1_{\rm{discrete}}(h_m,L) = \frac{1}{\sqrt{L}} \left( f^1 \left(
	\frac{h_m+a}{\sqrt{L}} \right)  
	+ \mathcal{O}\left(L^{-1}\right) \right)\;,
\end{equation}
which is a good trick to get rid of the leading finite size effects. Following this exact result for $\kappa =1$ (\ref{fss}), we thus
plot on Fig. \ref{fig_scaling_comp} (left) $\sqrt{L} P_{\rm{num}}^{\kappa}(h_m,L)$ as a function of $x=(h_m+a)/\sqrt{L}$ with $a = 1/\sqrt{2}$ for all $\kappa$ and observe a very good collapse of curves for different $L$.
%
 In Fig.~\ref{fig_scaling_comp} (right), we show 
a comparison between $\sqrt{L} P_{\rm{num}}^{\kappa}(h_m,L)$ and our analytical predictions in Eq. (\ref{distrib_real}) 
evaluated with Mathematica for different values of $\kappa = 0, 1/32, 1/2$ and $\kappa = 1$. The very good agreement between analytics and 
numerics is obtained here via a single ``fitting'' parameter $a = 1/\sqrt{2}$ for all values of $\kappa$, adjusted to take into account finite size corrections (\ref{fss}).

\section{Truncated area under brownian excursion}

In this last section we explore further the connections between the distribution of the MRH for arbitrary $\kappa$ and the distribution of areas below constrained Brownian motions \cite{janson_review}. Indeed, for $\kappa=1$, it was shown in Ref. \cite{satya_mrh,satya_mrh2}, that $f^1(x)$ describes the distribution of the area under a Brownian excursion on a unit interval.
\begin{figure}[h]
\begin{center}
\scalebox{.4}{\includegraphics[angle=0]{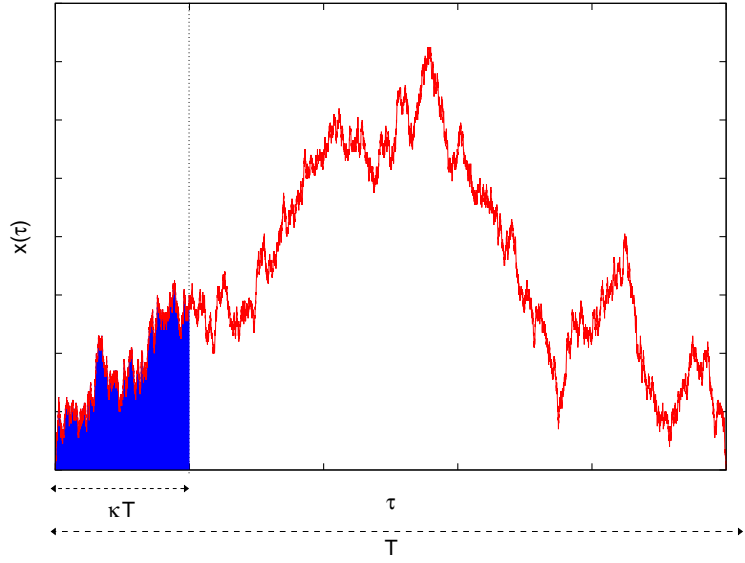}}
\caption{Truncated area (in blue) under a realization of a Brownian excursion (in red).}
\label{fig_area}
\end{center}
\end{figure}
Here we consider a Brownian excursion, {\it i.e.} a Brownian path $\{x(\tau),\ 0 \leq \tau \leq T\}$ starting and ending at the origin $x(0)=x(\tau)=0$ and which stays positive in-between $\forall \tau \in ]0,T[,\ x(\tau)>0$ and consider the random variable ${\cal A}_\kappa = \int_0^{\kappa L} x(\tau) d\tau$, which we call the {\it truncated area} (see Fig.~\ref{fig_area}). We compute its distribution, which can be obtained by a similar calculation as shown above. 

We start with the joint distribution of positions of the brownian excursion 
\begin{eqnarray}
P[x(\tau)]	& =\lim_{\epsilon \to 0} \frac{1}{{\cal Z}(\epsilon)} 
	e^{-\half \int_0^T (\partial_\tau x)^2 \rmd \tau} \delta(x(0)-x(T)) \delta(x(0)-\epsilon) 
	\prod_{0\leq \tau \leq T} \theta( x(\tau) ) \nonumber \\
		& =\lim_{\epsilon \to 0} \frac{1}{{\cal Z}(\epsilon)} 
	e^{-\int_0^T \rmd \tau \left\{ \half(\partial_\tau x)^2 + V_0(x(\tau))\right\} } 
	\delta(x(0)-x(T)) \delta(x(0)-\epsilon) \;,
\end{eqnarray}
where $\epsilon$ is a regularization parameter which is needed because of the non-smoothness of the Brownian path (see Ref.~\cite{satya_mrh,satya_mrh2}) and ${\cal Z}(\epsilon)$ is the normalization constant
\begin{equation}
{\cal Z}(\epsilon) = \int\limits_{x(0)=\epsilon}^{x(T)=\epsilon} \mathcal{D}x(\tau) 
e^{-\int_0^T \rmd \tau \left\{ \half (\partial_\tau x)^2 + V_0(x(\tau))\right\} } = G_0(\epsilon,T|\epsilon,0) \;,
\end{equation}
where $G_0(b,t_b|a,t_a)$ is the propagator of the free particle with a wall in $x=0$~(\ref{propagateur_02}). The distribution of the truncated area $\int_0^{\kappa T} x(\tau) \rmd \tau$, 
for the Brownian excursion $\{x(\tau),\ 0\leq \tau \leq T\}$ is then
\begin{equation}
P({\cal A}_{\kappa},T)	= \textrm{Prob} \left( \int_0^{\kappa T} x(\tau) \rmd \tau =A_{\kappa} \right)
				=\left< \delta \left( A_{\kappa}- \int_0^{\kappa T} x(\tau) \rmd \tau \right) \right> \;,
\end{equation}
where the mean value is taken respectively to the distribution of excursions $P[x(\tau)]$. We have
\begin{equation}
P({\cal A}_{\kappa},T) =\lim_{\epsilon \to 0}
	\frac{
	\int\limits_{x(0)=\epsilon}^{x(T)=\epsilon} \mathcal{D}x(\tau) 
	\delta \left( A_{\kappa}- \int_0^t x(\tau) \rmd \tau \right) \ 
	e^{-\int_0^T \rmd \tau \left\{ \half (\partial_\tau x)^2 + V_0(x(\tau))\right\} }
	}{
	\int\limits_{x(0)=\epsilon}^{x(T)=\epsilon} \mathcal{D}x(\tau) e^{-\int_0^T \rmd \tau \left\{ \half (\partial_\tau x)^2 + V_0(x(\tau))\right\} }
	} \;.
\end{equation}
We now take a Laplace transform with respect to ${\cal A}_{\kappa}$, which is a positive quantity. 
So we have
\begin{eqnarray}
\tilde{P}(p,T)	&=\int_0^\infty \rmd {\cal A}_{\kappa} \ P({\cal A}_{\kappa},T) \rme^{-p {\cal A}_{\kappa}}\\
			&=\lim_{\epsilon \to 0} \frac{
	\int\limits_{x(0)=\epsilon}^{x(T)=\epsilon} \mathcal{D}x(\tau) \ 
        	e^{-\int_0^T \rmd \tau \left\{ \half (\partial_\tau x)^2 + V_0(x(\tau))\right\} } \
	e^{-p \int_0^{\kappa T} \rmd \tau x(\tau)}
				}{
		G_0(\epsilon,T|\epsilon,0)
				}
\end{eqnarray}
By similar techniques, one can identify a convolution of propagators. 
Expanding numerator and denominator in powers of the regulator $\epsilon$,
one finds that the lowest order is $\epsilon^2$ for the numerator and the denominator.
Taking the limit $\epsilon \to 0$, one finds that the distribution of the truncated area obeys the following scaling form
\begin{equation}
P({\cal A}_{\kappa},T)=\frac{1}{(\kappa T)^{3/2}} 
	w^{\kappa}\left( \frac{{\cal A}_{\kappa}}{(\kappa T)^{3/2}} \right),
\end{equation}
with the scaling function, parameterized by $\kappa$, given by
\begin{eqnarray}
&&w^{\kappa}(x) = \frac{4 \sqrt{6}}{\pi} \, x^{-10/3} \int_0^{\infty} \rmd q \sum_{n=1}^{\infty} [q F_n(q)] \
\left( \frac{b_n(q,\kappa)}{\kappa} \right)^{2/3} \, 
\rme^{-b_n(q,\kappa)/(\kappa x^2)} \nonumber \\
&& U(-5/6,4/3,b_n(q,\kappa)/(\kappa x^2)) ,
\end{eqnarray}
with the same coefficient $b_n(q,\kappa)$ as in Eq. (\ref{distrib_real}) and $F_n(q)$ defined in Eq~\ref{Fnk}..
We can check easily that the scaling function $w^{\kappa}(x)$ 
converges to the Airy distribution function as $\kappa \to 1$, as shown in Refs~\cite{satya_mrh,satya_mrh2}. However, for $\kappa \neq 1$ these two distributions $f^{\kappa(x)}$ and  $w^{\kappa}(x)$ are different, albeit having similar expressions.

\section{Conclusion}

To conclude, we have introduced an alternative definition of the relative height $h^\kappa(x)$ of an elastic one-dimensional interface with pbc, indexed by
a real $\kappa \in [0,1]$, which interpolates smoothly between the height relative to the initial value when $\kappa \to 0$ and the height relative to the spatial average value for $\kappa \to 1$. We have obtained, using path-integral techniques, the exact distribution of the maximal relative height $f^\kappa(x)$ which interpolates between the well known Rayleigh ($\kappa = 0 $) and Airy ($\kappa = 1$) distribution (see Fig. \ref{plot_AiryRayleigh}). This thus constitutes one new family, parameterized by $\kappa$, of strongly correlated variables where such analytical calculations can be done. Although our calculations have been done for a continuum model (\ref{def_model}), the arguments presented in Ref. \cite{schehr_sos}, based on the Cental Limit Theorem, show that our results actually hold in the limit of large system size, for a wide class of lattice models of interfaces ${\cal H} = \sum_i |H_i - H_{i+1}|^p$ with arbitrary $p$ and in that sense this distribution $f^\kappa(x)$ in Eq. (\ref{distrib_real}) is {\it universal}. Finally we have shown that the method employed here to compute this distribution can be used to compute the distribution of the truncated area ${\cal A}_\kappa$ (see Fig.~\ref{fig_area}) under a Brownian excursion.

\appendix

\section{Quantum mechanics problems}

In our path integral calculations of distribution of MRH, we have to deal with simple one dimensional quantum mechanics problem.
This appendix summarizes the results used in the main text.

\subsection{Free particle}
\label{app_prop_free}

Let us consider a free particle in one dimension. The Hamiltonian is:
\begin{equation}
H_{\rm{free}}=-\half \frac{\rmd^2}{\rmd x^2} .
\end{equation}
The solution of the Schr\"odinger equation are the planes waves:
\begin{equation}
\varphi_k(x)= \frac{1}{\sqrt{2\pi}} \ e^{i k x},
\end{equation}
with $-\infty< k <+\infty$. The associated energy to $\varphi_k$ is $E_k=k^2/2$. The calculation of the propagator is done by decomposing onto the eigenbasis:
\begin{eqnarray}
G_{\rm{free}}(y,t|x,s)&= \langle y | \rme^{-(t-s) H_{\rm{free}}} | x \rangle \nn \\
	&= \int\limits_{-\infty}^{+\infty} \rmd k \ {\varphi_k}^*(y) \varphi_k(x) \rme^{-(t-s)k^2/2} \nn \\
\label{free_propagator}
	&= \frac{1}{\sqrt{2\pi (t-s)}} \exp \left( - \frac{(y-x)^2}{2(t-s)} \right)
\end{eqnarray}
Another way of doing it is to follow Feynman's prescription~\cite{feynman_hibbs}: 
if the lagrangian is a quadratic form of its variables, then the
propagator is proportional to the exponential of minus the classical action joining the two points in space-time.
Moreover, if the lagrangian does not depend explicitly on time, the constant of proportionality is simply a function of the difference
of the two times:
\begin{equation}
G(y,t|x,s)=F(t-s) \exp\left(- S[x_{\rm{classical}}(\tau)] \right) .
\end{equation}
When we are in imaginary time, because we are dealing directly with probabilities, the function $F(t-s)$ can be obtained by normalization.
For the example of the free particle, the action is:
\begin{equation}
S[x(\tau)]=\int\limits_s^t \rmd \tau \ \half \left( \frac{\rmd x(\tau)}{\rmd \tau} \right)^2,
\end{equation}
and the solution of the Euler-Lagrange equation between $(x,s)$ and $(y,t)$ is
\begin{equation}
x_{\rm{classical}}(\tau)= \frac{y-x}{t-s} (\tau-s) + x.
\end{equation}
Inserting it into the action, we recover directly the free propagator~(\ref{free_propagator}), by imposing the normalization
\begin{equation}
\int\limits_{-\infty}^{+\infty} \rmd y \ G(y,t|x,s) = 1 .
\end{equation}

\subsection{Free particle with a wall}
\label{app_prop_wall}

We impose that some region of space cannot be visited by the particle, for example, we want to keep our particle in the positive region $x>0$.
To do this, we put a potential
\begin{equation}
V_0(x)=
\cases{
+\infty &\rm{if $x<0$} \;,\\
0 & \rm{if $x>0$} \;.
} 
\end{equation}
The solution of the associated Schr\"odinger equation is
\begin{equation}
\varphi_k(x)=\sqrt{\frac{2}{\pi}} \sin (k x),
\end{equation}
with $0<k<+\infty$, for $x>0$, and $\varphi_k(x)=0$ for $x<0$. Now the normalization constant is found directly from what we impose on the free particle solution 
(the only freedom is the phase of the solution, who is unphysical). The associated energy is $k^2/2$.
\par

If the particle is constrained to stay below some value $M$, then
\begin{equation}
\varphi_k(x)=\sqrt{\frac{2}{\pi}} \sin (k(M- x)),
\end{equation}
with $0<k<+\infty$, for $x<M$ and $\varphi_k(x)=0$ for $x>M$.
\par
Let us compute the propagator when the wall is in $x=0$: 
\begin{eqnarray}
G_0(y,t|x,s) &= \int\limits_{0}^{+\infty} \rmd k \ \frac{2}{\pi} \sin (k y) \sin (k x) \rme^{- (t-s)k^2/2} \\
	&=\frac{1}{\sqrt{2 \pi (t-s)}} \left( \rme^{- \frac{(y-x)^2}{2(t-s)}} - \rme^{-\frac{(y+x)^2}{2(t-s)}} \right) \;.
\end{eqnarray}
The asymptotic form, when $t \to s$, of this propagator is just
\begin{equation}
\label{app_wallproplimit}
\lim_{t \to s } G_0(y,t|x,s) = \delta(y-x) - \delta(y+x) \;.
\end{equation}

\subsection{Particle in a box}
\label{app_prop_box}

The Hamiltonian is
\begin{equation}
H_{0,M}=-\half \frac{\rmd^2}{\rmd x^2} +V_{0,M}(x),
\end{equation}
with the box potential
\begin{equation}
V_{0,M}(x)=
\cases{
+\infty &\rm{if $x<0$} \;, \\
0 & \rm{if $0<x<M$} \;, \\
+\infty &\rm{if $x>M$} \;.
}
\end{equation}
The solution is indexed by a positive integer $n=1,2,\dots$ 
\begin{equation}
\psi_n(x)=\sqrt{\frac{2}{M}} \sin \left( \frac{n \pi}{M} x \right)
\end{equation}
with energy $E_n=n^2\pi^2/(2M^2)$ and the propagator is
\begin{equation}
G_{0,M}(y,t|x,s)=\sum_{n=1}^{+\infty} \frac{2}{M} \sin\left( \frac{n \pi}{M} y \right) \sin \left( \frac{n \pi}{M} x \right) \exp \left(- (t-s) \frac{n^2 \pi^2}{2 M^2} \right) .
\end{equation}

\subsection{Linear potential and a wall: Airy potential}
\label{app_prop_airy}

The name `Airy potential' is used for conveniency, and like in the main text, we use in this subsection the index `Airy' for all quantity referred to
the quantum mechanical problem of a particle in the potential
\begin{equation}
V_{\rm{Airy}}(x) =
\cases{
+\infty & \rm{if $-\infty < x < 0$} , \\
\lambda \, x & \rm{if $0 < x < +\infty$} \;,
}
\end{equation}
where $\lambda$ is a real positive parameter. We want to solve the stationary Schr\"odinger equation
\begin{equation}
H_{\rm{Airy}} \psi_n(x) = -\half \frac{\rmd^2 \psi_n}{\rmd x^2}(x) + V_{\rm{Airy}}(x) \, \psi_n(x) = E_n \, \psi_n(x) ,
\end{equation}
Because of the confining potential, we already know that the energy spectrum will be discrete, thus we use an integer $n$ index.
For $x>0$, one can re-write this equation 
\begin{equation}
\psi_n''(x)-2 \lambda \left( x - \frac{E_n}{\lambda} \right) \psi_n(x) = 0 .
\end{equation}
Putting $\psi_n(x)=\chi(x-E_n/\lambda)$, then $\chi(y)=\varphi((2\lambda)^{1/3} y)$, one obtains 
\begin{equation}
\varphi''(u)-u \,\varphi(u) = 0 ,
\end{equation}
which is solved by the Airy functions ($\textrm{Ai}(x)$ and $\textrm{Bi}(x)$). That explains the name `Airy' used since the beginning.
We eliminate $\textrm{Bi}(x)$ because the wave function need to tend to zero when $x\to \infty$.
The solutions are of the form $\psi_n(x) = (1/K) \ind{Ai}((2\lambda)^{1/3} (x - E_n/\lambda) $ with $K$ a normalization constant.
The energy levels are determined by the condition that the wave function vanishes in $x=0$. They are then relied to the zeros $-\alpha_n$
of the Airy function $\textrm{Ai}(x)$ on the negative axis ($\alpha_n>0$):
\begin{equation}
\ind{Ai} \left( - \frac{2^{1/3}}{\lambda^{2/3}} \, E_n \right) = 0  \quad
\Rightarrow \quad E_n = \alpha_n \,2^{-1/3} \lambda^{2/3} \quad (n=1,2,\dots).
\end{equation}
The normalized wave functions are:
\begin{eqnarray}
\psi_n(x)& =\frac{\ind{Ai}\left[(2\lambda)^{1/3} x - \alpha_n \right] }{ \sqrt{\int_0^{\infty} \ind{Ai}^2 \left[(2\lambda)^{1/3} y - \alpha_n \right] \rmd y}} \nn \\
&= \frac{\ind{Ai}\left[(2\lambda)^{1/3} x - \alpha_n \right] }{\sqrt{(2\lambda)^{-1/3} \left( \ind{Ai}'[-\alpha_n] \right)^2 }} \;,
\end{eqnarray}
where from the first to the second line, we have just modified the expression of the normalization constant, 
using the property that $\int_{-\alpha_n}^{\infty} \ind{Ai}^2(z) \rmd z = \left( \ind{Ai}'(-\alpha_n) \right)^2$. Let us compute the propagator of the particle in such a potential.
\begin{eqnarray}
&&\hspace*{-2.3cm}G_{\rm{Airy}}(y,t|x,s) = \langle y | \rme^{-(t-s) H_{\ind{Airy}} } | x \rangle  \\
&& = \sum_{n=1}^{\infty} {\psi_n}^*(y) \psi_n(x)\, \rme^{-(t-s) E_n} \nn \\
&& = \sum_{n=1}^{\infty} \frac{\ind{Ai}\left[(2\lambda)^{1/3} y - \alpha_n \right]}{\sqrt{(2\lambda)^{-1/3} \left( \ind{Ai}'[-\alpha_n] \right)^2} } \ 
\frac{\ind{Ai}\left[(2\lambda)^{1/3} x - \alpha_n \right] }{\sqrt{(2\lambda)^{-1/3} \left( \ind{Ai}'[-\alpha_n] \right)^2}}  \
\rme^{-(t-s) \alpha_n 2^{-1/3} \lambda^{2/3} }  \nn \;.
\end{eqnarray}
The general form of the Airy propagator, for the potential $V_{\rm{Airy}}(x)=V_0(x)+\lambda x$, is
\begin{eqnarray}
\label{airy_formule1}
&&\hspace*{-2.5cm}G_{\rm{Airy}}(y,t|x,s) = (2 \lambda)^{1/3} \sum_{n=1}^{\infty} 
\frac{\ind{Ai}\left[(2\lambda)^{1/3} y - \alpha_n \right] \, \ind{Ai}\left[(2\lambda)^{1/3} x - \alpha_n \right] }
{\left(\ind{Ai}'[-\alpha_n] \right)^2} \ \rme^{-\alpha_n (t-s) 2^{-1/3} \lambda^{2/3} } . \nn \\
\end{eqnarray}

\subsection{Linear potential and asymptotic form of the Airy propagator}
\label{app_prop_lin}

For the purpose of the main text, we have to obtain the asymptotic form of the Airy propagator in Eq. (\ref{airy_formule1}) when $|t-s| \to 0$. Here, we consider the simple case of the linear potential (constant force), we compute exactly the propagator
and then we take the limit, showing that this is equivalent to taking the limit of the Airy problem.
First, the action reads, between $(x,s)$ and $(y,t)$ along the path $x(x)$:
\begin{equation}
S[x(\tau)]=\int_{s}^{t} \rmd \tau \left\{ \half \left( \frac{\rmd x(\tau)}{\rmd \tau} \right)^2 + \lambda x(\tau) \right\} .
\end{equation}
Because the Lagrangian is a quadratic form in its variables $(x(\tau),\dot{x}(\tau))$, we know that the propagator is of the form
\begin{equation}
G_{\ind{lin}}(y,t|x,s)= F(t-s)\,\rme^{-S[x_{\rm{classical}}(\tau)]},
\end{equation}
where $x_{\rm{classical}}(\tau)$ is the classical path between $(x,s)$ and $(y,t)$.
Without loss of generality because of time translation invariance, let us take $s=0$. The classical path is
\begin{equation}
x_{\rm{classical}}(\tau) = \half \lambda \tau^2 + \left\{ (y-x)-\half \lambda t^2 \right\} \frac{\tau}{t} + x .
\end{equation}
The action, computed along the classical path, is then
\begin{equation}
S[x_{\rm{classical}}(\tau)] = -\frac{1}{24} \lambda^2 t^3 + \frac{(y-x)^2}{2t} + \half \lambda t (x+y)
\end{equation}
Because we are dealing directly with probabilities, we can compute the function $F(t)$ by the normalization condition
\begin{equation}
\int_{-\infty}^{+\infty} \rmd y\ G_{\ind{lin}}(y,t|x,0) = 1 .
\end{equation}
Thus we obtain
\begin{equation}
\hspace*{-2cm}G_{\ind{lin}}(y,t|x,s)= \frac{1}{\sqrt{2 \pi (t-s)}} \exp\left( - \frac{\lambda (t-s)^3}{8} - \frac{(y-x)^2}{2(t-s)} - \frac{\lambda (t-s)}{2} (x+y) \right) .
\end{equation}
For an infinitesimal difference in time, \textit{i.e.} for $t=s+\epsilon$, we have
\begin{equation}
\label{prop_lin_1}
G_{\ind{lin}}(y,s+\epsilon|x,s)= \frac{1}{\sqrt{2 \pi \epsilon}} \exp\left( - \frac{\lambda^2 \epsilon^3}{8} - \frac{(y-x)^2}{2\epsilon} - \frac{\lambda \epsilon}{2} (x+y) \right) .
\end{equation}
Here, let's remark that $\lambda$ can have a dependence in the time difference $\epsilon$ (which can have any value at this stage).
That's not introducing any dangerous time-dependence: we give a linear potential $\lambda (\epsilon) x$ for the particle at point $x$,
with a slope given by any function of $\epsilon$, and we ask what is the probability of finding at certain point $y$ at time $t=s+\epsilon$,
given that the particle started at point $x$ at time $s$. However, we see in the previous formula that the expression 
of the propagator becomes ill-defined when $\epsilon \to 0$ as soon as $\lambda(\epsilon) \propto \epsilon^{-(1+\alpha)}$, 
with $\alpha>0$. 
\par
Now let us take $\lambda=p/\epsilon$, with $p$ a fixed positive number. We have at leading order in $\epsilon \ll 1$
\begin{equation}
G_{\ind{lin}}(y,s+\epsilon|x,s)= \frac{1}{\sqrt{2 \pi \epsilon}} \exp\left( - \frac{(y-x)^2}{2\epsilon} - \frac{p}{2} (x+y) \right)  
+ \mathcal{O}(\epsilon^{1/2}).
\end{equation} 
The limit $\epsilon \to 0$ can be taken in formula~(\ref{prop_lin_1}), and one obtains 
\begin{equation}
\label{app_airyproplimit}
\lim_{\epsilon \to 0} G_{\ind{lin}}(y,s+\epsilon|x,s)= \delta(y-x) \ \rme^{-p x}.
\end{equation}

The difference between the linear propagator and the Airy propagator is that there is in the latter a wall in $x=0$. 
But if we have to take the Airy propagator between two infinitesimally close points,
the particle, placed in $x>0$ does not have the time to feel the wall. Indeed, for a time $\epsilon$, 
the Brownian particle will explore a space region of order $\epsilon^{1/2}$.
So, as close as we are from the wall, we can choose an $\epsilon$ small enough not to feel the wall. In this limit, 
the Airy propagator is equivalent to the linear propagator, that is, for $t=s+\epsilon$,
\begin{equation}
\lim_{\epsilon \to 0} G_{\rm{Airy}}(y,s+\epsilon|x,s) = \lim_{\epsilon \to 0} G_{\ind{lin}}(y,s+\epsilon|x,s) ,
\end{equation}
for the same expression for the linear part of the potential.
With the potential $V_{\rm{Airy}}(x)=V_0(x)+\frac{p}{\epsilon} x$, we thus have
\begin{equation}
\lim_{\epsilon \to 0} G_{\rm{Airy}}(y,s+\epsilon|x,s) =  \delta(y-x) \ \rme^{-p x}.
\end{equation}
Explicitely, this yields the identity
\begin{eqnarray}
&&\hspace*{-2.5cm}\label{identity_airy}
\lim_{\epsilon \to 0} \left( \frac{2 p}{\epsilon} \right)^{1/3}
\sum_{n=1}^{\infty} \frac{ \textrm{Ai}\left[ \left( \frac{2 p}{\epsilon} \right)^{1/3} y - \alpha_n \right] 
\textrm{Ai}\left[ \left( \frac{2 p}{\epsilon} \right)^{1/3} x - \alpha_n \right] }
{\left(\textrm{Ai}'[-\alpha_n]\right)^2}
e^{-\alpha_n 2^{-1/3} \epsilon^{1/3} p^{2/3}}
= \delta(y-x) \ \rme^{-p x} \;. \nonumber \\
\end{eqnarray}

\section{Another calculation of the normalization constant $\tilde Z_\kappa$}


In this paragraph, we compute the normalization constant $\tilde Z_\kappa$ involved in Section~\ref{h_kappa} 
using the Random Acceleration Process, 
which is described by the stochastic equation of motion $\ddot x(t) = \zeta(t)$ where
$\zeta$ is Gaussian white noise. 
From the joint distribution of heights $ P[\{h^\kappa\}] $, one can
determine the marginal distribution of one height, say
$h^\kappa(x_0)$, for a particular $x_0$. The joint distribution of the
heights is Gaussian and therefore the marginal distribution of a
single height must be Gaussian (by virtue of the central limit
theorem). 
One simple way to compute $\tilde Z_\kappa$ is to consider the marginal
distribution of $h^\kappa(\kappa L)$ : let us call 
$p(v)=\textrm{Prob}(h^\kappa(\kappa L)=v)$ this probability. $p(v)$ is
a centered Gaussian with a variance $\sigma^2$ which we now determine.

Integrating the joint distribution $P[\{h^\kappa\}]$ over all paths
(bridges, {\it i.e.} with pbc) such that 
$ h^\kappa(0)=h^\kappa(L)=u$ for $u \in
]-\infty,\infty[$, and keeping $h^\kappa(\kappa L)=v$ fixed, one recovers $p(v)$~: 
\begin{equation}\label{expr_pv} 
\hspace*{-1cm}p(v) = \int_{-\infty}^{+\infty} \rmd u
\int\limits_{h^\kappa(0)=u}^{h^\kappa(L)=u} \mathcal{D}h^\kappa(x) 
P[\{h^\kappa\}]  \delta(h^\kappa(\kappa L)-v) = 
\frac{1}{\sqrt{2 \pi \sigma^2}} \, \rme^{- v^2/(2 \sigma^2)} .
\end{equation}

The path integral in this last expression (\ref{expr_pv}) reads explicitly: 
\begin{eqnarray}
&&p(v)  = \frac{1}{\tilde Z_\kappa} \int_{-\infty}^{+\infty} du \int\limits_{h^\kappa(0)=u}^{h^\kappa(L)=u}
              \mathcal{D}h^\kappa(x) 
                e^{-\half \left[ \int_0^{\kappa L} d x
               (\partial_{x} h^\kappa)^2 + \int_{\kappa L}^L d x
              (\partial_{x} h^\kappa)^2 \right] } \nn \\
&&\times \delta\left(\int_0^{\kappa L} h^\kappa(x) d x \right) \,
             \delta(h^\kappa(\kappa L) - v)  \nn \;.
\label{interpol_normalization_calculus}
\end{eqnarray}
As $h^\kappa(\kappa L)=v$ is fixed, and because the process is
Markovian, one can cut the whole path integral into two independent
blocks~: one for the time interval $[0,\kappa L]$ and one for the
interval $[\kappa L,L]$. This yields
\begin{eqnarray}
&&p(v) = \frac{1}{\tilde Z_\kappa}\int_{-\infty}^{+\infty} \rmd u  
                \int\limits_{h^\kappa(\kappa L)=v}^{h^\kappa(L)=u} \mathcal{D}h^\kappa(x)
                e^{-\half \int_{\kappa L}^{L} \rmd x (\partial_{x} h^\kappa)^2 } \nn \\
&&\times \ \int\limits_{h^\kappa(0)=u}^{h^\kappa(\kappa L)=v} \mathcal{D}h^\kappa(x) \
                e^{-\half \int_0^{\kappa L} \rmd x (\partial_{x} h^\kappa)^2 }
                \ \delta\left(\int_0^{\kappa L} h^\kappa(x) \rmd x\right) 
\end{eqnarray}
The block for $[\kappa L,L]$ is simply the propagator of the free
Brownian motion, but the path integral for $[0,\kappa L]$ contains an
additional delta function, which constraint the Brownian trajectories
to have a null area. In the latter, let us write $r(x)=\int_0^x h^\kappa(x') \rmd x'$, so that $ h^\kappa(x) = 
(\rmd r(x) / \rmd x) = \dot{r}(x)$, and 
\begin{eqnarray}
&&\int\limits_{h^\kappa(0)=u}^{h^\kappa(\kappa L)=v} \mathcal{D}h^\kappa(x) \
                e^{-\half \int_0^{\kappa L} \rmd x (\partial_{x} h^\kappa)^2 }
                \ \delta\left(\int_0^{\kappa L} h^\kappa(x) \rmd x\right) \nn \\
&&\equiv 
\int\limits_{\dot{r}(0)=u}^{\dot{r}(\kappa L)=v} \mathcal{D}\dot{r}(x) \
		e^{-\half \int_0^{\kappa L} \rmd x (\partial_{x} \dot{r}(x))^2 }
                \ \delta\left( r(\kappa L) \right) .
\end{eqnarray}
We recognize the propagator of the Random Acceleration Process (RAP)
between position 
$r(0)=0$ and speed $\dot r(0)=u$ and position $r(\kappa L)=0$ and
speed $\dot{r}(\kappa L)=v$. 
The propagator $G_{\rm{RAP}}$ is the probability of finding the randomly accelerated particle 
at point $x_2$ with speed $v_2$ at time $t=T$, knowing that it was in point $x_1$ with speed $v_1$ at time $t=0$,
and is given by the formula~\cite{burkhardt_review}
\begin{eqnarray}
\label{prop_rap}
\hspace*{-2.5cm}G_{\rm{RAP}} (x_2,v_2,T|x_1,v_1,0) = \frac{\sqrt{3}}{\pi T^2} \, 
\exp \left( -\frac{6}{T^3} (x_2 - x_1 - v_1 T) (x_2 - x_1 - v_2 T) - \frac{2}{T} (v_2-v_1)^2 \right) \nn \\
\end{eqnarray}
Hence
\begin{equation}
p(v) = \frac{1}{\tilde Z_\kappa} \int\limits_{-\infty}^{+\infty} \rmd u \left\{
G_{\rm{free}}(u,L|v,\kappa L) \ \times \ G_{\ind{RAP}}(0,v,\kappa L|0,u,0) \right\} .
\end{equation}
Using formula~(\ref{prop_rap}), one is reduced to
compute a gaussian integral, and one finds 
\begin{equation}
p(v) = \frac{1}{\tilde Z_\kappa} \, \frac{1}{\pi} \, \sqrt{\frac{3}{\kappa^3
  L^4(4-3\kappa)}} \ e^{-\frac{6}{\kappa L(4-3\kappa)} v^2} .
\end{equation}
We can finally identify the second moment of the marginal distribution 
\begin{equation}
\sigma^2  = \langle (h^\kappa(\kappa L))^2 \rangle = \frac{L(4-3\kappa)}{12} ,
\end{equation}
and the normalization constant
\begin{equation}\label{al}
\tilde Z_\kappa = \frac{1}{\kappa L \sqrt{2\pi L} } ,
\end{equation}
which is what we found in the main text using a different method (cf.~Eq.~(\ref{zkappa})).


\end{document}